\documentclass[preprint]{elsarticle}
\usepackage[english]{babel}
\usepackage{verbatim}
\usepackage{alltt}
\usepackage{graphicx}
\usepackage{multirow}
\usepackage{varwidth}
\usepackage[centering,includeheadfoot,margin=3.4cm]{geometry}

\newcommand{\etexttt}[1]{\textit{#1}}

\makeatletter
\g@addto@macro\@verbatim\footnotesize
\makeatother

\newcommand{\sepv}{\vspace{-6.0mm}}

\begin{document}

\title{The ModelCC Model-Based Parser Generator}

\author{Luis~Quesada, Fernando~Berzal, and Juan-Carlos~Cubero\\
  Department of Computer Science and Artificial Intelligence, \\
  CITIC, University of Granada, \\ 
  Granada 18071, Spain \\
  \textit{\{lquesada, fberzal, jc.cubero\}@decsai.ugr.es}
  }

\begin{abstract}

Formal languages let us define the textual representation of data with precision.
Formal grammars, typically in the form of BNF-like productions, describe the language syntax, which is then annotated for syntax-directed translation and completed with semantic actions.
When, apart from the textual representation of data, an explicit representation of the corresponding data structure is required, the language designer has to devise the mapping between the suitable data model and its proper language specification, and then develop the conversion procedure from the parse tree to the data model instance.
Unfortunately, whenever the format of the textual representation has to be modified, changes have to propagated throughout the entire language processor tool chain.
These updates are time-consuming, tedious, and error-prone.
Besides, in case different applications use the same language, several copies of the same language specification have to be maintained.
In this paper, we introduce ModelCC, a model-based parser generator that decouples language specification from language processing, hence avoiding many of the problems caused by grammar-driven parsers and parser generators.
ModelCC incorporates reference resolution within the parsing process.
Therefore, instead of returning mere abstract syntax trees, ModelCC is able to obtain abstract syntax graphs from input strings.


\end{abstract}

\maketitle

\section{Introduction}

A formal language represents a set of strings \cite{Jurafsky2009}.
Formal languages consist of an alphabet, which describes the basic symbol or character set of the language, and a grammar, which describes how to write valid sentences of the language.
In Computer Science, formal languages are used, among other things, for the precise definition of data formats and the syntax of programming languages.
The front-end of a language processor, such as an interpreter or compiler, determines the grammatical structure corresponding to the textual representation of data conforming to a given language specification.
Such grammatical structure is typically represented in the form of a parse tree.

Most existing language specification techniques \cite{Aho1972} require the language designer to provide a textual specification of the language grammar.
The proper specification of such a grammar is a nontrivial process that depends on the lexical and syntactic analysis techniques to be used, since each kind of technique requires the grammar to comply with different constraints.
Each analysis technique is characterized by its expression power and this expression power determines whether a given analysis technique is suitable for a particular language.
The most significant constraints on formal language specification originate from the need to consider context-sensitivity, the need of performing an efficient analysis, and some techniques' inability to consider grammar ambiguities or resolve conflicts caused by them.

In its most general sense, a model is anything used in any way to represent something else.
In such sense, a grammar is a model of the language it defines.

In Software Engineering, data models are also common.
Data models explicitly determine the structure of data.
Roughly speaking, they describe the elements they represent and the relationships existing among them.
From a formal point of view, it should be noted that data models and grammar-based language specifications are not equivalent, even though both of them can be used to represent data structures.
A data model can express relationships a grammar-based language specification cannot.
Moreover, a data model does not need to comply with the constraints a grammar-based language specification has to comply with.
Hence describing a data model is generally easier than describing the corresponding grammar-based language specification.

When both a data model and a grammar-based language processor are required, the traditional implementation of the language processor requires the language designer to build a grammar-based language specification from the data model and also to implement the conversion from parse trees to data model instances.

Whenever the language specification has to be modified, the language designer has to manually propagate changes throughout the entire language processor tool chain, from the specification of the grammar defining the formal language (and its adaptation to specific parsing tools) to the corresponding data model.
These updates are time-consuming, tedious, and error-prone.
By making such changes labor-intensive, the traditional approach hampers the maintainability and evolution of the language \cite{Kats2010}.

Moreover, it is not uncommon that different applications use the same language.
For example, the compiler, different code generators, and the editor or the debugger within an IDE typically need to grapple with the full syntax of a programming language.
Their maintenance typically requires keeping several copies of the same language specification synchronized.

As an alternative approach, a language can also be defined by a data model that, in conjunction with the declarative specification of some constraints, can be automatically converted into a grammar-based language specification \cite{Quesada2011c}.

This way, the data model representing the language can be modified as needed, without having to worry about the language processor and the peculiarities of the chosen parsing technique, since the corresponding language processor will be automatically updated.

Furthermore, as the data model is the direct representation of a data structure, such data structure can be implemented as an abstract data type (in object-oriented languages, as a set of collaborating classes).
Following the proper software design principles, that implementation can be performed without having to embed or mix semantic actions with the language specification, as it is typically the case with grammar-driven language processors.

Finally, as the data model is not bound to a particular parsing technique, evaluating alternative and/or complementary parsing techniques is possible without having to propagate their constraints into the language model.
Therefore, by using an annotated data model, model-based language specification completely decouples language specification from language processing, which can be performed using whichever parsing techniques that are suitable for the formal language implicitly defined by the model.

It should be noted that, while, in general, the result of the parsing process is an abstract syntax tree that corresponds to a valid parsing of the input text according to the language concrete syntax, nothing prevents the model-based language designer from modeling non-tree structures.
Indeed, a model-driven parser generator can automate, among other syntactic and semantic checks that are typically deferred to later stages in the language processing pipeline, the implementation of reference resolution mechanisms \cite{Quesada2012k,Quesada2014a}.

In this paper, we introduce ModelCC, a model-based tool for language specification.
As a parser generator that decouples language specification from language processing, it avoids many of the problems caused by grammar-driven parsers and parser generators.
ModelCC is able to resolve references and obtain abstract syntax graphs as the result of the parsing process, rather than the traditional abstract syntax trees obtained from conventional parser generators.

The structure of this paper is as follows.
Section \ref{sec:background} introduces formal grammars and surveys parsing algorithms and tools.
Section \ref{sec:modelbased} presents the philosophy behind model-based language specification.
Section \ref{sec:modelspecification} comments on ModelCC building blocks.
Section \ref{sec:modelconstraints} describes the model constraints supported by ModelCC, which declaratively define the features of the formal language defined by the model.
Section \ref{sec:example1} shows a prototypical example, which is used to discuss the advantages of model-based language specification over traditional grammar-based language specification.
Section \ref{sec:example2} shows a complex example, which illustrates abstract syntax graph parsing via the resolution of references between language elements.
Lastly, Section \ref{sec:conclusionsfuturework} presents our conclusions and the future work that derives from our research.

\section{Background} \label{sec:background}

In this section, we introduce formal grammars (Subsection \ref{subsec:formalgrammars}), describe the typical architecture of language processor front-ends (Subsection \ref{subsec:architecture}), survey key parsing algorithms (Subsection \ref{subsec:algorithms}), review existing lexer and parser generators (Subsection \ref{subsec:generators}), and describe language workbenches (Subsection \ref{subsec:workbenches}).

\subsection{Formal Grammars} \label{subsec:formalgrammars}

Formal grammars are used to specify the syntax of a language \cite{Ginsburg1975,Harrison1978}.
A grammar naturally describes the hierarchical structure of language constructs \cite{Aho2006}.
Using a set of rules, a grammar describes how to form strings from the language alphabet that are valid according to the language syntax.
A grammar $G$ is formally defined \cite{Chomsky1956} as the tuple $(N,\Sigma,P,S)$, where:

\begin{itemize}
\item $N$ is the finite set of nonterminal symbols of the language, sometimes called syntactic variables, none of which appear in the language strings.
\item $\Sigma$ is the finite set of terminal symbols of the language, also called tokens, which constitute the language alphabet (i.e. they appear in the language strings).
      Therefore, $\Sigma$ is disjoint from $N$.
\item $P$ is a finite set of productions, each one of the form $(\Sigma \cup N)^{*} N (\Sigma \cup N)^{*} \rightarrow (\Sigma \cup N)^{*}$, where $*$ is the Kleene star operator, $\cup$ denotes set union, the part before the arrow is called the left-hand side of the production, and the part after the arrow is called the right-hand side of the production.
\item $S$ is a distinguished nonterminal symbol, $S \in N$: the grammar start symbol.
\end{itemize}

For convenience, when several productions share their left-hand side, they can be grouped into a single production containing their the shared left-hand side and all their different right-hand sides separated by $|$.

Context-free grammars are formal grammars in which the left-hand side of each production consists of only a single nonterminal symbol.
All their productions, therefore, are of the form $N \rightarrow (\Sigma \cup N)^{*}$. Context-free grammars generate context-free languages.

A context-free grammar is said to be ambiguous if there exists at least one string that can be generated by the grammar in more than one way.
In fact, some context-free languages are inherently ambiguous (i.e. all context-free grammars generating them are ambiguous).

An attribute grammar is a formal way to define attributes for the symbols in the productions of a formal grammar, associating values to these attributes.
Semantic rules annotate attribute grammars and define the value of an attribute in terms of the values of other attributes and constants \cite{Aho2006}.
The key feature of attribute grammars is that they let us transfer information from anywhere in the abstract syntax tree to anywhere else in a controlled and formal way, hence their frequent use in syntax-directed translation.

When using attribute grammars, language specification often becomes low-level when dealing with non-local dependencies, that is, when an attribute of a syntax tree node is dependent on attributes of nodes far away in the syntax tree \cite{boyland:98fiber}.
A reference attribute grammar is a formalism that allows explicit references denoting nodes arbitrarily far away in the syntax tree and to access these attributes via the references \cite{Hedin1987}.
Reference attribute grammars allow the specification of complex static semantic analysis constraints in an object-oriented form.

Graph grammars \cite{Ehrig1999} allow the manipulation of graphs based on productions: if the left-hand side of a production matches the working graph or a subgraph of it, it can be replaced with the right-hand side of the production.
These grammars can be used to define the syntax of visual languages.

\subsection{The Architecture of Language Processors} \label{subsec:architecture}

The architecture of a language-processing system decomposes language processing into several steps which are typically grouped into two phases: analysis and synthesis.
The analysis phase, which is responsibility of the language processor front end, starts by breaking up its input into its constituent pieces (lexical analysis or scanning) and imposing a grammatical structure upon them (syntax analysis or parsing).
The language processor back end will later synthesize the desired target from the results provided by the front end.

A lexical analyzer, also called lexer or scanner, processes an input string conforming to a language specification and produces the tokens found within it.
Lexical ambiguities occur when a given input string simultaneously corresponds to several token sequences \cite{Nawrocki1991}, in which tokens may overlap.

A syntactic analyzer, also called parser, processes sequences of input tokens and determines their grammatical structure with respect to the given language grammar, usually in the form of parse trees.
In the absence of lexical ambiguities, the parser input consists of a stream of tokens, whereas it will be a directed acyclic graph of tokens when lexical ambiguities are present.
Syntactic ambiguities occur when a given set of tokens simultaneously corresponds to several parse trees \cite{Aho1975}.

\subsection{Scanning and Parsing Algorithms} \label{subsec:algorithms}

Scanning and parsing algorithms are characterized by the expression power of the languages to which they can be applied, their support for ambiguities or lack thereof, and the constraints they impose on language specifications.

Traditional lexers are based on a finite-state machine that is built from a set of regular expressions \cite{McNaughton1960}, each of which describes a token type.
The efficiency of regular expression lexers is $O(n)$, being $n$ the input string length.

Lamb \cite{Quesada2011a} is a lexer with lexical ambiguity support that allows the specification of tokens not only by regular expressions, but also by arbitrary pattern matching classes.
Lamb also supports token type precedences.
The efficiency of the Lamb lexer is, in the worst case, $O(n^2t^2)$, being $n$ the input string length and $t$ the number of different token types.
In practical cases, in which lexical ambiguities seldom arise, the efficiency of the Lamb lexer is close to $O(n)$, being $n$ the input string length.
Lamb is able to recognize ambiguities that involve tokens of different types that overlap, in contrast to traditional approaches such as the Schr{\"o}dinger's token \cite{Aycock2001}, which cannot be applied to overlapping tokens as it only allows for strings to have a superposition of token types.

Efficient parsers for certain classes of context-free grammars exist.
These include top-down LL parsers, which construct a leftmost derivation of the input sentence, and bottom-up LR parsers, which construct a rightmost derivation of the input sentence.

LL grammars were formally introduced in \cite{Lewis1968}, albeit LL(k) parsers predate their name \cite{Oettinger1961}.
An LL parser is called an LL(k) parser if it uses $k$ lookahead tokens when parsing a sentence, while it is an LL(*) parser if it is not restricted to a finite set of $k$ lookahead tokens and it can make parsing decisions by recognizing whether the following tokens belong to a regular language \cite{Jarzabek1975,Nijholt1976}, or by using syntactic or semantic predicates \cite{Parr1995}.
While LL(k) parsers are always linear, LL(*) ranges from $O(n)$ to $O(n^2)$.

LR parsers were introduced by Knuth \cite{Knuth1965}.
DeRemer later developed the LALR \cite{DeRemer1969,DeRemer1982} and SLR \cite{DeRemer1971} parsers that are in use today.

Efficient LR and LL parsers for certain classes of ambiguous grammars are also possible by using simple disambiguating rules \cite{Aho1975,Earley1975}.

When parsing theory was originally developed, machine resources were scarce, and so parser efficiency was the paramount concern \cite{Parr2011}.
Hence all the aforementioned parsing algorithms parse in linear time (i.e. their efficiency is $O(n)$, being $n$ the input string length) and they do not support syntactic ambiguities.
Free from the requirement to develop efficient linear-time parsing algorithms, researchers have developed many powerful nondeterministic parsing strategies following both the top-down approach (LL parsers) and the bottom-up approach (LR parsers).

Following the top-down approach, Packrat parsers \cite{Ford2002packrat} and their associated Parsing Expression Grammars (PEGs) \cite{Ford2004peg} preclude only the use of left-recursive grammar rules.
Even though they use backtracking, packrat parsers are linear rather than exponential because they memoize partial results, they attempt the alternative productions in the specified order, and they accept only the first one that matches an input position.
In fact, LL(*) is an optimization of packrat parsing \cite{Parr2011}.

A general-purpose dynamic programming algorithm for top-down parsing context-free grammars was independently developed by Cocke \cite{Cocke1970}, Younger \cite{Younger1967}, and Kasami \cite{Kasami1965}: the CYK parser.
This general-purpose algorithm is $O(n^3)$ for ambiguous and unambiguous context-free grammars.
The Earley parser \cite{Earley1970} is another general-purpose dynamic programming algorithm for top-down parsing context-free grammars that executes in cubic time ($O(n^3)$) in the general case, quadratic time ($O(n^2)$) for unambiguous grammars, and linear time ($O(n)$) for almost all LR(k) grammars.
In both cases, the result is a parse forest with all possible interpretations of the input.

Following the bottom-up approach, Generalized LR (GLR) is an extension of LR parsers that handles nondeterministic and ambiguous grammars.
GLR forks new subparsers to pursue all possible actions emanating from nondeterministic LR states, terminating any subparsers that lead to invalid parses.
The result is, again, a parse forest with all possible interpretations of the input.
GLR parsers perform in linear to cubic time, depending on how closely the grammar conforms to the underlying LR strategy.
The time required to run the algorithm is proportional to the degree of nondeterminism in the grammar.
Bernard Lang is typically credited with the original GLR idea \cite{Lang1974}.
Later, Tomita used the algorithm for natural language processing \cite{Tomita1985}.
Tomita's Universal parser \cite{Tomita1987}, however, failed for grammars with epsilon rules (i.e. productions with an empty right-hand side).
Several extensions have been proposed that support epsilon rules \cite{Farshi1991,Rekers1992,Ishii1994,McPeak2004}.
GLR parsing is an extension of an LR parser that handles nondeterministic and ambiguous grammars.

The Fence parser \cite{Quesada2012f} is an optimized Earley-like algorithm that supports ambiguous context-free grammars, epsilon productions, and constraints on associativity, composition, and precedence, as well as custom constraints.
Hence, while in theory the Fence parser is as efficient as Earley, CYK, or GLR parsers, in the practice it is more efficient, since defined constraints can render subparsings invalid on the fly.
The Fence parser also supports recursive productions by halting infinite loops as soon as a production cycle has been performed on the same input text.

\subsection{Lexer and Parser Generators} \label{subsec:generators}

Lexer and parser generators are tools that take a language specification as input and produce a lexer or parser as output.
They can be characterized by their input syntax, their ability to specify semantic actions, and the parsing algorithms the resulting parsers implement.

Lex \cite{lex} and yacc \cite{yacc} are commonly used in conjunction \cite{Levine1992}.
They are the default lexer generator and parser generator, respectively, in many Unix environments and standard compiler textbooks (e.g. \cite{Aho2006}) often use them as examples.
Lex is the prototypical regular-expression-based lexer generator, while yacc and its many derivatives generate LALR parsers.

JavaCC \cite{McManis1996} is a parser generator that creates LL(k) parsers, albeit it has been superseded by ANTLR \cite{Parr1995}.
ANTLR is a parser generator that creates LL(*) parsers.
ANTLR-generated parsers are linear in practice and greatly reduce speculation, reducing the memoization overhead of pure packrat parsers.
Furthermore, ANTLR supports tree parsing, in which instead of symbol sequences, abstract syntax trees are processed.

The Rats! \cite{Grimm2006} packrat parser generator is a PEG-based tool that also optimizes memoization to improve its speed and reduce its memory footprint.
Like ANTLR, it does not accept left-recursive grammars.
Unlike ANTLR, programmers do not have to deal with conflict messages, since PEGs have no concept of a grammar conflict: they always choose the first possible interpretation.
This can lead to unexpected behavior in the case of ambiguous grammars, which may pass unnoticed.

NLyacc \cite{Ishii1994} and Elkhound \cite{McPeak2004} are examples of GLR parser generators. Elkhound achieves yacc-like parsing speeds when grammars are LALR(1).
Like PEG parsers, GLR parsers silently accept ambiguous grammars and programmers have to detect ambiguities dynamically \cite{Parr2011}.

YAJco \cite{Poruban2009} is an interesting tool that accepts, as input, a set of Java classes with annotations that specify the prefixes, suffixes, operators, tokens, parentheses, and optional elements common in typical programming languages.
As output, YAJco generates a BNF-like grammar specification for JavaCC \cite{McManis1996}.
Since YAJco is built on top of a parser generator, the language designer has to be careful when annotating his classes, as the implicit grammar he is defining has to comply with the constraints imposed by the underlying LL(k) parser generator.

\subsection{Language Workbenches} \label{subsec:workbenches}

Language workbenches are tools that allow the definition of new languages which are fully integrated with each other and whose primary source of information is a persistent abstract representation \cite{language-workbenches}.

During the 1980s and 1990s, compiler- and IDE-generating systems were often built around abstract syntax models such as attribute grammars.
These systems are powerful tools that do not only specify compilers.

The Synthesizer Generator \cite{Reps1989} accepts high-order attribute grammar specifications as input and generates syntax editors.
The LRC system \cite{Kuiper1998} is based on the same input language used by the Synthesizer Generator, but generates programming environments with graphical interfaces and abstract semantic tree navigators.

VL-Eli \cite{Kastens2002} is an extension to the Eli system \cite{Kastens1998} that generates visual languages, just as VisPro \cite{Zhang1998} does.

Other interesting features of some language workbenches are pretty printing \cite{Swierstra1998} and program animations \cite{Saraiva2002}.

FNC-2 \cite{Jourdan2002} is an attribute-grammar-based system that generates, apart from the scanner and parser pair and a pretty printer, an incremental attribute evaluator and a dependency graph visualizer.

The CENTAUR system \cite{Borras1988} accepts formal operational semantics specifications as input and generates, apart from the scanner and parser pair, a pretty printer, a syntax-directed editor, a type checker, an interpreter, and graphic tools.
It was later extended into the SmartTools system \cite{Attali2001}, which also generates a compiler, a structured editor, and XML related tools.

The ASF+SDF environment \cite{Brand2001} takes algebraic specifications as input and generates a scanner and parser pair, a pretty printer, a syntax-directed editor, a type checker, an interpreter, and a debugger.

The Gem-Mex system \cite{Anlauff1998} resorts to abstract state machines for language specification and generates a scanner and parser pair, a type checker, an interpreter, and a debugger.

The Mj\o lner BETA metaprogramming system \cite{Machura1994} provides the SbyS editor, which supports the user in editing language constructs defined by the language grammar.
The APPLAB language laboratory \cite{Bjarnason1996} extends this concept further by allowing the grammar to be edited by the very same SbyS editor and by supporting multiple views of the same abstract model.

The Spoofax/IMP workbench \cite{Kats2010}, which is based on the Stratego program transformation language \cite{Bravenboer2008}, uses SDF for describing grammars and produces DSL implementations that can be embedded in Java-based enterprise systems, and Rascal.
Both Spoofax and the entry-level Simpl workbench \cite{Freudenthal2013} use the IMP toolkit for creating Eclipse-based IDEs.

The Rascal metaprogramming language \cite{Bos2011} is a domain-specific language and a toolset for source code analysis and manipulation.
Similarly to Spoofax, it uses the SDF language for describing the concrete syntax of the DSL.

OMeta \cite{Warth2007} is an object-oriented language for pattern matching that provides a straightforward way for programmers to implement scanners, parsers, visitors, and tree transformers, which can be extended using familiar object-oriented mechanisms.

PECAN \cite{Reiss1984} is a generator of program development systems for algebraic programming languages that produces complete environments from an abstract syntax tree.
The language developer does not have access to the abstract syntax tree, but instead modifies it through views or concrete representations such as a syntax-directed editor, a Nassi-Schneiderman structured flowchart, a declaration editor, or a module interconnection diagram.

The LISA system \cite{Henriques2005} advocates an 'attribute grammar = class' paradigm that enables incremental language development, and is able to produce editors, inspectors, debuggers, and animators.

The Kermeta language workbench \cite{Jezequel2013} involves a EMOF-like \cite{EMOF} meta-language for abstract syntax specification, an OCL-like \cite{OCL} meta-language for static semantics, and a meta-language for behavioral semantics.

\section{Model-Based Language Specification} \label{sec:modelbased}

In this section, we discuss the concepts of abstract and concrete syntax (Subsection \ref{subsec:asmcsm}), analyze the potential advantages of model-based language specification (Subsection \ref{subsec:modelbased}), and compare our proposed approach with the traditional grammar-driven language design process (Subsection \ref{subsec:comparison}).

\subsection{Abstract Syntax and Concrete Syntaxes} \label{subsec:asmcsm}

The abstract syntax of a language is just a representation of the structure of the different elements of a language without the superfluous details related to its particular textual representation \cite{Kleppe2007}.
On the other hand, a concrete syntax is a particularization of the abstract syntax that defines, with precision, a specific textual or graphical representation of the language.
It should also be noted that a single abstract syntax can be shared by several concrete syntaxes \cite{Kleppe2007}.

For example, the abstract syntax of the typical \emph{$<$if$>$-$<$then$>$-$<$optional else$>$} statement in imperative programming languages could be described as the concatenation of a conditional expression and one or two statements.
Different concrete syntaxes could be defined for such an abstract syntax, which would correspond to different textual representations of a conditional statement, e.g. \{``if'', ``('', expression, ``)'', statement, optional ``else'' followed by another statement\} and \{``IF'', expression, ``THEN'', statement, optional ``ELSE'' followed by another statement, ``ENDIF''\}.

The idea behind model-based language specification is that, starting from a single abstract syntax model (ASM) representing the core concepts in a language, language designers would later develop one or several concrete syntax models (CSMs).
These concrete syntax models would suit the specific needs of the desired textual or graphical representation for the language sentences.
The ASM-CSM mapping could be performed, for instance, by annotating the abstract syntax model with the constraints needed to transform the elements in the abstract syntax into their concrete representation.

\subsection{Advantages of Model-Based Language Specification} \label{subsec:modelbased}

Focusing on the abstract syntax of a language offers some benefits \cite{Kleppe2007} and provides some potential advantages to model-based language specification over the traditional grammar-based language specification approach:

\begin{itemize}

\item
When reasoning about the features a language should include, specifying its abstract syntax seems to be a better starting point than working on its concrete syntax details.
In fact, we control complexity by building abstractions that hide details when appropriate \cite{sicp}.

\item
Sometimes, different incarnations of the same abstract syntax might be better suited for different purposes (e.g. an human-friendly syntax for manual coding, a machine-oriented format for automatic code generation, a Fit-like \cite{fit} syntax for testing, different architectural views for discussions with project stakeholders...).
Therefore, it might be useful for a given language to support multiple syntaxes.

\item
Since model-based language specification is independent from specific lexical and syntactic analysis techniques, the constraints imposed by specific parsing algorithms do not affect the language design process.
In principle, however, it might not be even necessary for the language designer to have advanced knowledge on parser generators when following a model-driven language specification approach.

\item
A full-blown model-driven language workbench \cite{language-workbenches,Reiss1985,Borras1988,Hedin1988,Bjarnason1996,Grundy2013} would allow the modification of a language abstract syntax model and the automatic generation of a working IDE on the run.
The specification of domain-specific languages would become easier, as the language designer could play with the language specification and obtain a fully-functioning language processor on the fly, without having to worry about the propagation of changes throughout the complete language processor tool chain.
\end{itemize}

In short, the model-driven language specification approach brings domain-driven design \cite{ddd} to the domain of language design.
It provides the necessary infrastructure for what Evans would call the `supple design' of language processing tools: the intention-revealing specification of languages by means of abstract syntax models, the separation of concerns in the design of language processing tools by means of declarative ASM-CSM mappings, and the automation of a significant part of the language processor implementation.

\subsection{Comparison with the Traditional Approach} \label{subsec:comparison}

\begin{figure}[tb!]
\centering
\includegraphics[scale=0.24]{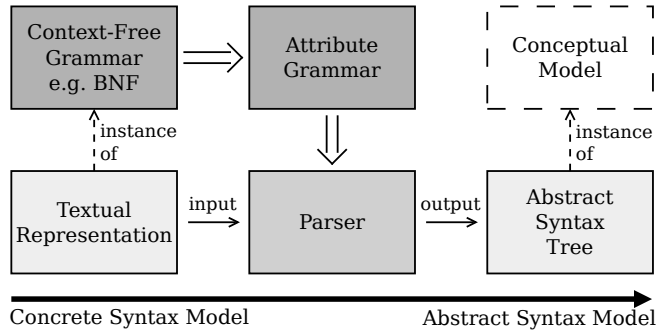}
\caption{Traditional language processing approach.} \label{fig:traditional}
\end{figure}

\begin{figure}[tb!]
\centering
\includegraphics[scale=0.24]{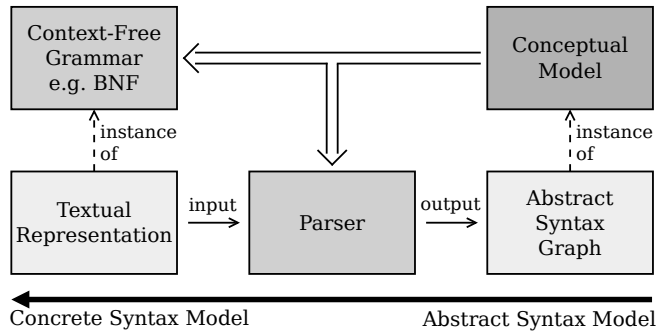}
\caption{Model-based language processing approach.} \label{fig:ModelCC}
\end{figure}

A diagram summarizing the traditional language design process is shown in Figure \ref{fig:traditional}, whereas the corresponding diagram for
the model-based approach proposed in this paper is shown in Figure \ref{fig:ModelCC}.

When following the traditional grammar-driven approach, the language designer starts by designing the grammar corresponding to the concrete syntax of the desired language, typically in BNF or a similar format.
Then, the designer annotates the grammar with attributes and, probably, semantic actions, so that the resulting attribute grammar can be fed into lexer and parser generator tools that produce the corresponding lexer and parser, respectively.
The resulting syntax-directed translation process generates abstract syntax trees from the textual representation in the concrete syntax of the language.

When following the model-driven approach, the language designer starts by designing the conceptual model that represents the abstract syntax of the desired language, focusing on the elements the language will represent and their relationships.
Instead of dealing with the syntactic details of the language from the start, the designer devises a conceptual model for it (i.e. the abstract syntax model, or ASM), the same way a database designer starts with an implementation-independent conceptual database schema before he converts that schema into a logical schema that can be implemented in the particular kind of DBMS that will host the final database.
In the model-driven language design process, the ASM would play the role of entity-relationship diagrams in database design and each particular CSM would correspond to the final table layout of the physical database schema in a relational DBMS.

Even though the abstract syntax model of the language could be converted into a suitable concrete syntax model automatically, the language designer will often be interested in specifying the details of the ASM-CSM mapping.
With the help of constraints imposed over the abstract model, the designer will be able to guide the conversion from the ASM to its concrete representation using a particular CSM.
This concrete model, when it corresponds to a textual representation of the abstract model, will be described by a formal grammar.
It should be noted, however, that the specification of the ASM is independent from the peculiarities of the desired CSM, as a database designer does not consider foreign keys when designing the conceptual schema of a database.
Therefore, the grammar specification constraints enforced by particular parsing tools will not impose limits on the design of the ASM.
The model-driven language processing tool will take charge of that and, ideally, it will employ the most efficient parsing technique that works for the language resulting from the ASM-CSM mapping.

While the traditional language designer specifies the grammar for the concrete syntax of the language, annotates it for syntax-directed processing, and obtains an abstract syntax tree that is an instance of the implicit conceptual model defined by the grammar, the model-based language designer starts with an explicit full-fledged conceptual model and specifies the necessary constraints for the ASM-CSM mapping.
In both cases, parser generators create the tools that parse the input text in its concrete syntax.
The difference lies in the specification of the grammar that drives the parsing process, which is hand-crafted in the traditional approach and automatically-generated as a result of the ASM-CSM mapping in the model-driven process.

Another difference stems from the fact that the result of the parsing process is an instance of an implicit model in the grammar-driven approach while that model is explicit in the model-driven approach.
An explicit conceptual model is absent in the traditional language design process albeit that does not mean that it does not exist.
On the other hand, the model-driven approach enforces the existence of an explicit conceptual model, which lets the proposed approach reap the benefits of domain-driven design.

There is a third difference between the grammar-driven and the model-driven approaches to language specification.
While, in general, the result of the parsing process is an abstract syntax tree that corresponds to a valid parsing of the input text according to the language concrete syntax, nothing prevents the conceptual model designer from modeling non-tree structures, which describe grammars with a power of expression similar to reference attribute grammars \cite{Burger2010}.
Hence the use of the `abstract syntax graph' term in Figure \ref{fig:ModelCC}.
This might be useful, for instance, for modeling graphical languages, which are not constrained by the linear nature of the traditional syntax-driven specification of text-based languages.

Instead of going from a concrete syntax model to an implicit abstract syntax model, as it is typically done, the model-based language specification process goes from the abstract to the concrete.
This alternative approach facilitates the proper design and implementation of language processing systems by decoupling language processing from language specification, which is now performed by imposing declarative constraints on the ASM-CSM mapping.

\section{ModelCC Model Specification} \label{sec:modelspecification}

Once we have described model-driven language specification in general terms, we now proceed to introduce ModelCC, a tool that supports our proposed approach to the design of language processing systems.
ModelCC, at its core, acts as a parser generator.
The starting abstract syntax model is created by defining classes that represent language elements and establishing relationships among those elements (associations in UML terms).
Once the abstract syntax model is established, its incarnation as a concrete syntax is guided by the constraints imposed over language elements and their relationships as annotations on the abstract syntax model.
In other words, the declarative specification of constraints over the ASM establishes the desired ASM-CSM mapping.

In this section, we introduce the basic constructs that allow the specification of abstract syntax models, while we will discuss how model constraints help us establish a particular ASM-CSM mapping in the following section of this paper.
Basically, the ASM is built on top of basic language elements, which might be viewed as the tokens in the model-driven specification of a language.
Model-driven language processing tools such as ModelCC provide the necessary mechanisms to combine those basic elements into more complex language constructs, which correspond to the use of concatenation, selection, and repetition in the syntax-driven specification of languages.

Our final goal is to allow the specification of languages in the form of abstract syntax models such as the one shown in Figure \ref{fig:calcmodelcc}, which will be used as an example in Section \ref{sec:example1}.
This model, in UML format, specifies the abstract syntax model of the language supported by a simple arithmetic expression language.
The annotations that accompany the model provide the necessary information for establishing the complete ASM-CSM mapping that corresponds to the traditional infix notation for arithmetic expressions.
Moreover, the model also incorporates the method that lets us evaluate such arithmetic expressions.
Therefore, Figure \ref{fig:calcmodelcc} represents a complete interpreter for arithmetic expressions in infix notation using ModelCC (its implementation as a set of cooperating Java classes appears in Figures \ref{fig:calcimmodelcc1} and \ref{fig:calcimmodelcc2}.

As mentioned above, the specification of the ASM in ModelCC starts with the definition of basic language elements, which can be modeled as simple classes in an object-oriented programming language.
The ASM-CSM mapping of those basic elements will establish their correspondence to the tokens that appear in the concrete syntax of the language whose ASM we design in ModelCC.

In the following subsections, we describe the mechanisms provided by ModelCC to implement the three main constructs that let us specify complete abstract syntax models on top of basic
language elements.

\subsection{Concatenation}

\begin{figure}[tb!]
\centering
\includegraphics[scale=1]{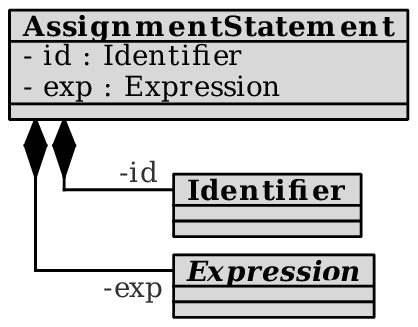}
\caption{An assignment statement as an example of element composition (concatenation in textual CSM terms).} \label{fig:assignmentstatement}
\end{figure}

Concatenation is the most basic construct we can use to combine sets of language elements into more complex language elements.
In textual languages, this is achieved just by joining the strings representing its constituent language elements into a longer string that represents the composite language element.

In ModelCC, concatenation is achieved by object composition.
The resulting language element is the composite element and its members are the language elements the composite element collates.

When translating the ASM into a textual CSM, each composite element in a ModelCC model generates a production rule in the grammar representing the CSM.
This production, with the nonterminal symbol of the composite element in its left-hand side, concatenates the nonterminal symbols corresponding to the constituent elements of the composite element in its right-hand side.
By default, the order of the constituent elements in the production rule is given by the order in which they are specified in the object composition, but such an order is not strictly necessary (e.g. many ambiguous languages might require differently ordered sequences of constituent elements and even some unambiguous languages allow for unordered sequences of constituent elements).

The model in Figure \ref{fig:assignmentstatement} shows an example of object composition in ASM terms that corresponds to string concatenation in CSM terms.
In this example, an assignment statement is composed of an identifier, i.e. a reference to its l-value, and an expression, which provides its r-value.
In a textual CSM, the composite \emph{AssignmentStatement} element would be translated into the following production rule: \etexttt{$<$AssignmentStatement$>$ ::= $<$Identifier$>$ $<$Expression$>$}.
Obviously, such production would probably include some syntactic sugar in an actual programming language, either for avoiding potential ambiguities or just for improving its readability and writability, but that is the responsibility of ASM-CSM mappings, which will be analyzed in Section \ref{sec:modelconstraints}.

\subsection{Selection}

Selection is necessary as a language modeling primitive operation to represent choices, so that we can specify alternative elements in language constructs.

In ModelCC, selection is achieved by subtyping.
Specifying inheritance relationships among language elements in an object-oriented context corresponds to defining `is-a' relationships in a more traditional database design setting.
The language element we wish to establish alternatives for is the superelement (i.e. the superclass in OO or the supertype in DB modeling), whereas the different alternatives are represented as subelements (i.e. subclasses in OO, subtypes in DB modeling).
Alternative elements are always kept separate to enhance the modularity of ModelCC abstract syntax models and their integration in language processing systems.

In the current version of ModelCC, multiple inheritance is not supported, albeit the same results can be easily simulated by combining inheritance and composition.
We can define subelements for the different inheritance hierarchies representing choices so that those subelements are composed by the single element that appears as a common choice in the different scenarios.
This solution fits well with most existing programming languages, which do not always support multiple inheritance, and avoids the pollution of the shared element interface in the ASM, which would appear as a side effect of allowing multiple inheritance in abstract syntax models.

\begin{figure}[tb!]
\centering
\includegraphics[scale=1]{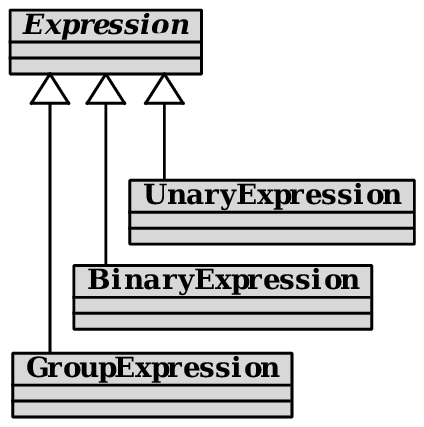}
\caption{Subtyping for representing choices in ModelCC.} \label{fig:expression}
\end{figure}

Each inheritance relationship in ModelCC, when converting the ASM into a textual CSM, generates a production rule in the CSM grammar.
In those productions, the nonterminal symbol corresponding to the superelement appears in its left-hand side, while the nonterminal symbol of the subelement appears as the only symbol in the production right-hand side.
Obviously, if a given superelement has $k$ different subelements, $k$ different productions will be generated representing the $k$ alternatives defined by the abstract syntax model.

The model shown in Figure \ref{fig:expression} illustrates how an arithmetic \emph{Expression} can be either an \emph{UnaryExpression}, a \emph{BinaryExpression}, or a \emph{GroupExpression} in the language defined for a simple calculator.
The grammar resulting from the conversion of this ASM into a textual CSM would be: \etexttt{$<$Expression$>$ ::= $<$UnaryExpression$>$ $|$ $<$BinaryExpression$>$ $|$
$<$GroupExpression$>$}.

\subsection{Repetition}

Representing repetition is also necessary in abstract syntax models, since a language element might appear several times in a given language construct.
When a variable number of repetitions is allowed, mere concatenation does not suffice.

Repetition is also achieved though object composition in ModelCC, just by allowing different multiplicities in the associations that connect composite elements to their constituent elements.
The cardinality constraints described in Section \ref{sec:modelconstraints} can be used to annotate ModelCC models in order to establish specific multiplicities for repeatable language elements.

Each composition relationship representing a repetitive structure in the ASM will lead to two additional production rules in the grammar defining a textual CSM: a recursive production of the form \etexttt{$<$ElementList$>$ ::= $<$Element$>$ $<$ElementList$>$} and a complementary production \etexttt{$<$ElementList$>$ ::= $<$Element$>$}, where \etexttt{$<$Element$>$} is the nonterminal symbol associated to the repeating element.
Obviously, an equivalent non-left-recursive derivation could also be obtained if needed (e.g. when generating a specific parser that does not support left-recursive rules).

It should also be noted that \etexttt{$<$ElementList$>$} will take the place of the nonterminal \etexttt{$<$Element$>$} in the production derived from the composition relationship that connects the repeating element with its composite element (see the above section on how composition is employed to represent concatenation in ModelCC).

In practice, repeating elements will often appear separated in the concrete syntax of a textual language, hence repeatable elements can be annotated with separators, as we will see in Section \ref{sec:modelconstraints}.
In case separators are employed, the recursive production derived from repeatable elements will be of the form \etexttt{$<$ElementList$>$ ::= $<$Element$>$ $<$Separator$>$ $<$ElementList$>$}.

When a repeatable language element is optional, i.e. its multiplicity can be $0$, an additional epsilon production is appended to the grammar defining the textual CSM derived from the ASM: \etexttt{$<$ElementList$>$ ::= $\epsilon$}.

\begin{figure}[tb!]
\centering
\includegraphics[scale=1]{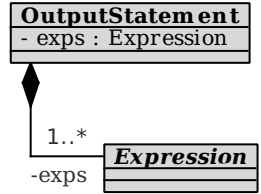}
\caption{Multiple composition for representing repetition in ModelCC.} \label{fig:output}
\end{figure}

For example, the model in Figure \ref{fig:output} shows that an \emph{OutputStatement} can include several \emph{Expression}s, which will be evaluated for their results in order for them to be sent to the corresponding output stream.
This ASM would result in the following textual CSM grammar: \etexttt{$<$OutputStatement$>$ ::= $<$ExpressionList$>$} for describing the composition and \etexttt{$<$ExpressionList$>$ ::= $<$Expression$>$ $<$ExpressionList$>$ $|$ $<$Expression$>$} for allowing repetition.

\section{ModelCC Model Constraints} \label{sec:modelconstraints}

Once we have examined the mechanisms that let us create abstract syntax models in ModelCC, we now proceed to describe how constraints can be imposed on such models in order to establish the desired ASM-CSM mapping.
As soon as that ASM-CSM mapping is established, ModelCC is able to generate the suitable parser for the concrete syntax defined by the CSM.

ModelCC allows the definition of constraints using metadata annotations or a domain-specific language.

Now supported by all the major programming platforms, metadata annotations have been used in reflective programming and code generation \cite{Fowler2002}.
Among many other things, they can be employed for dynamically extending the features of your software development runtime \cite{Berzal2005} or even for building complete model-driven software development tools that benefit from the infrastructure provided by your compiler and its associated tools \cite{mdsd-ideal}.

The ModelCC domain-specific language for ASM-CSM mappings \cite{Quesada2014b} supports the separation of concerns in the design of language processing tools by allowing the definition of different CSMs for a common ASM.

In ModelCC, which supports language composition without being scannerless, a first set of constraints is used for pattern specification, a necessary feature for defining the lexical elements of the concrete syntax model, i.e. its tokens.

A second set of constraints is employed for defining delimiters in the concrete syntax model, whose use is common for eliminating language ambiguities or just as syntactic sugar in many languages.

A third set of ModelCC constraints lets us impose cardinalities on language elements, which control element repeatability and optionality.

A fourth set of constraints lets us impose evaluation order on language elements, which are employed to declaratively resolve further ambiguities in the concrete syntax of a textual language by establishing associativity, precedence, and composition policies, the latter employed, for example, for resolving the ambiguities that cause the typical shift-reduce conflicts in LR parsers.

A fifth set of constraints lets us specify the element constituent order in composite elements.

A sixth set of constraints lets us specify referenceable language elements and references to them.

Finally, custom constraints let us provide specific lexical, syntactic, and semantic constraints that take into consideration context information.

Table \ref{fig:tablesummary} summarizes the set of constraints supported by ModelCC for establishing ASM-CSM mappings between abstract syntax models and their concrete representation in textual CSMs.

In the following sections, we provide working examples of languages implemented using ModelCC.

\begin{table*}[tb]
\begin{center}

\setlength{\tabcolsep}{7pt}
\resizebox{\linewidth}{!}{
\begin{tabular}{ l  l  l } \hline
\normalsize
Constraints on & Annotation & Function \\ \hline

\multirow{2}{*}{... patterns}
& @Pattern & Pattern matching specification of basic language elements. \\
& @Value & Field where the recognized input token will be stored. \\ \hline

\multirow{3}{*}{... delimiters}
& @Prefix & Element prefix(es). \\
& @Suffix & Element suffix(es). \\
& @Separator & Element separator(s) in lists of elements. \\ \hline

\multirow{2}{*}{... cardinality}
& @Optional & Optional elements.\\
& @Multiplicity & Minimum and maximum element multiplicity.\\ \hline

\multirow{3}{*}{... evaluation order}
& @Associativity & Element associativity (e.g. left-to-right). \\
& @Composition & Eager or lazy composition for nested composites. \\
& @Priority & Element precedence level/relationships. \\ \hline

\multirow{2}{*}{... composition order}
& @Position & Element member relative position. \\
& @FreeOrder & When there is no predefined order among element members. \\ \hline

\multirow{2}{*}{... references}
& @ID & Identifier of a language element. \\
& @Reference & Reference to a language element. \\ \hline

\multirow{2}{*}{Custom constraints}
& \multirow{2}{*}{@Constraint} & \multirow{2}{*}{Custom user-defined constraint.} \\ \\ \hline
\end{tabular}
}
\end{center}
\caption{The constraints supported by the ModelCC model-based parser generator.} \label{fig:tablesummary}
\end{table*}

\section{A Working Example} \label{sec:example1}

We proceed to compare how an interpreter for arithmetic expressions can be implemented using conventional tools and how it can be implemented using the model-driven approach using ModelCC.
Albeit the arithmetic expression example in this section is necessarily simplistic, it already provides some hints on the potential benefits model-driven language specification can bring to more challenging endeavors.
This example language is also used in the next section as a base for a more complex language, which illustrates ModelCC capabilities of language composition.

First, we will outline the features we wish to include in our simple arithmetic language. Later, we will describe how an interpreter for this language is built using two of the most established tools in use by language designers: lex \& yacc on the one hand, ANTLR on the other.
Finally, we will implement the same language processor using ModelCC by defining an abstract syntax model.
This ASM will be annotated to specify the required ASM-CSM mapping and it will also include the necessary logic for evaluating arithmetic expressions.
This example will let us compare ModelCC against conventional parser generators and it will be used for discussing the potential advantages provided by our model-driven language specification approach.

\subsection{Language Description}

Our language will employ classical arithmetic expressions in infix notation with the following capabilities:

\begin{itemize}

\item Unary operators: +, and -.
\item Binary operators: +, -, *, and /, being - and / left-associative.
\item Operator priorities: * and / precede + and -.
\item Group expressions, which are delimited by parentheses.
\item Integer and floating-point literals.

\end{itemize}

\subsection{Conventional Implementation}

Using conventional tools, the language designer would start by specifying the grammar defining the arithmetic expression language in a BNF-like notation.
The BNF grammar shown in in Figure \ref{fig:calctrad} meets the requirements of our simple language, albeit it is not yet suitable for being used with existing parser generators, since they impose specific constraints on the format of the grammar depending on the parsing algorithms they employ.

\begin{figure*}[tb!]
\begin{verbatim}
<Expression> ::= <GroupExpression> | <BinaryExpression> | <UnaryExpression> | <LiteralExpression>
<GroupExpression> ::= '(' <Expression> ')'
<BinaryExpression> ::= <Expression> <BinaryOperator> <Expression>
<UnaryExpression> ::= <UnaryOperator> <Expression>
<LiteralExpression> ::= <RealLiteral> | <IntegerLiteral>
<BinaryOperator> ::= '+' | '-' | '/' | '*'
<UnaryOperator> ::= '+' | '-'
<RealLiteral> ::= <IntegerLiteral> '.' | <IntegerLiteral> '.' <IntegerLiteral>
<IntegerLiteral> ::= <Digit> <IntegerLiteral> | <Digit>
<Digit> ::= '0' | '1' |'2' | '3' | '4' | '5' | '6' | '7' | '8' | '9'
\end{verbatim}
\caption{BNF grammar for the arithmetic expression language.} \label{fig:calctrad}
\end{figure*}

\subsubsection{Lex \& yacc implementation}

When using lex \& yacc, the language designer converts the BNF grammar into a grammar suitable for LR parsing.
A suitable lex/yacc implementation defining the arithmetic expression grammar is shown in Figures \ref{fig:calclex} and \ref{fig:calcyacc}.

\begin{figure*}[tb!]
\begin{verbatim}
%{
#include "y.tab.h"
extern YYSTYPE yylval;
%}
%%
[0-9]+\.[0.9]*  return RealLiteral;
[0-9]+          return IntegerLiteral;
\+|\-           return UnaryOrPriority2BinaryOperator;
\/|\*           return Priority1BinaryOperator;
\(              return LeftParenthesis;
\)              return RightParenthesis;
.               ;
%%
\end{verbatim}
\caption{lex specification of the arithmetic expression language.} \label{fig:calclex}
\end{figure*}
\begin{figure*}[tb!]
\begin{verbatim}
%left UnaryOrPriority2BinaryOperator
%left Priority1BinaryOperator
%token IntegerLiteral RealLiteral LeftParenthesis RightParenthesis
%start Expression
%%
Expression : RealLiteral
           | IntegerLiteral
           | LeftParenthesis Expression RightParenthesis
           | UnaryOrPriority2BinaryOperator Expression
           | Expression UnaryOrPriority2BinaryOperator Expression
           | Expression Priority1BinaryOperator Expression
           ;
%%
#include "lex.yy.c"
int main(int argc,char *argv[]) { yyparse(); }
int yyerror(char *s) { printf("%s",s); }
\end{verbatim}
\caption{yacc specification of the arithmetic expression language.} \label{fig:calcyacc}
\end{figure*}

\begin{figure*}[tb!]
\begin{verbatim}
%{ #include "string.h"
   #include "y.tab.h"
   extern YYSTYPE yylval; %}
%%
[0-9]+\.[0.9]*  { yylval.value = atof(yytext); return RealLiteral; }
[0-9]+          { yylval.value = (double)atoi(yytext); return IntegerLiteral; }
\+|\-           {
                  if (yytext[0] == '+') yylval.operator = PLUSORADDITION;
                  else /*yytext[0] == '-'*/ yylval.operator = MINUSORSUBTRACTION;
                  return UnaryOrPriority2BinaryOperator;
                }
\/|\*           {
                  if (yytext[0] == '*') yylval.operator = MULTIPLICATION;
                  else /*yytext[0] == '/'*/ yylval.operator = DIVISION;
                  return Priority1BinaryOperator;
                }
\(              { return LeftParenthesis; }
\)              { return RightParenthesis; }
\n              { return LineReturn; }
.               ;
%%
\end{verbatim}
\caption{Complete lex implementation of the arithmetic expression interpreter.} \label{fig:calcimlex}
\end{figure*}

\begin{figure*}[tb!]
\begin{verbatim}
%left UnaryOrPriority2BinaryOperator
%left Priority1BinaryOperator
%token IntegerLiteral RealLiteral LeftParenthesis RightParenthesis
%start Line
%{
#include <stdio.h>
#define YYSTYPE attributes
typedef enum { PLUSORADDITION, MINUSORSUBTRACTION, MULTIPLICATION, DIVISION } optype;
typedef struct {
  optype operator;
  double value;
} attributes;
%}
%%
Expression : RealLiteral       { $$.value = $1.value; }
           | IntegerLiteral    { $$.value = $1.value; }
           | LeftParenthesis Expression RightParenthesis  { $$.value = $2.value; }
           | UnaryOrPriority2BinaryOperator Expression {
               if ($1.operator == PLUSORADDITION) $$.value = $2.value;
               else /*$1.operator == MINUSORSUBTRACTION*/ $$.value = -$2.value;
             }
           | Expression UnaryOrPriority2BinaryOperator Expression {
               if ($2.operator == PLUSORADDITION) $$.value = $1.value+$3.value;
               else /*$2.operator == MINUSORSUBTRACTION*/ $$.value = $1.value-$3.value;
             }
           | Expression Priority1BinaryOperator Expression {
               if ($2.operator == MULTIPLICATION) $$.value = $1.value*$3.value;
               else /*$2.operator == DIVISION*/ $$.value = $1.value/$3.value;
             }
           ;
Line       : Expression LineReturn { printf("%f\n",$1.value); } ;
%%
#include "lex.yy.c"
int main(int argc,char *argv[]) { yyparse(); }
int yyerror(char *s) { printf("%s",s); }
\end{verbatim}
\caption{Complete yacc implementation of the arithmetic expression interpreter.} \label{fig:calcimyacc}
\end{figure*}

Since lex does not support lexical ambiguities, the \emph{UnaryOperator} and \emph{BinaryOperator} nonterminals from the BNF grammar in Figure \ref{fig:calctrad} have to be refactored in order to avoid the ambiguities introduced by the use of + and - both as unary and binary operators.
A typical solution consists of creating the \emph{UnaryOrPriority2BinaryOperator} token type for representing them and then adjusting the grammar accordingly.
This token will act as an \emph{UnaryOperator} in \emph{UnaryExpression}s, and as a \emph{BinaryOperator} in \emph{BinaryExpression}s.

A similar solution is necessary for distinguishing different operator priorities, hence different token types are defined for each precedence level in the language, even though they perform the same role from a conceptual point of view.
The order in which they are declared in the yacc specification determines their relative priorities (please, note that these declarations are also employed to define operator associativity).

Unfortunately, the requirement to resolve ambiguities by refactoring the grammar defining the language involves the introduction of a certain degree of duplication in the language specification: separate token types in the lexer and multiple parallel production rules in the parser.

Once all ambiguities have been resolved, the language designer completes the lex \& yacc introducing semantic actions to perform the necessary operations.
In this case, albeit somewhat verbose in C syntax, the implementation of an arithmetic expression evaluator is relatively straightforward using the yacc \$ notation, as shown in Figures \ref{fig:calcimlex} and \ref{fig:calcimyacc}.
In our particular implementation of the arithmetic expression interpreter, carriage returns are employed to output results, hence our use of the ancillary \emph{Line} token type and \emph{LineReturn} nonterminal symbol.

\subsubsection{ANTLR implementation}

When using ANTLR, the language designer converts the BNF grammar into a grammar suitable for LL parsing.
An ANTLR specification of our arithmetic expression language is shown in Figure \ref{fig:calcantlr}.

\begin{figure*}[tb!]
\begin{verbatim}
grammar ExpressionEvaluator;
expression1 : expression2 ( '+' expression1 | '-' expression1 )* ;
expression2 : expression3 ( '*' expression2 | '/' expression2 )* ;
expression3 : '(' expression1 ')'
            | '+' expression1
            | '-' expression1
            | INTEGERLITERAL
            | FLOATLITERAL
            ;
INTEGERLITERAL : '0'..'9'+ ;
FLOATLITERAL : ('0'..'9')+ '.' ('0'..'9')* ;
NEWLINE : '\r'? '\n' ;
\end{verbatim}
\caption{ANTLR specification of the arithmetic expression language.} \label{fig:calcantlr}
\end{figure*}
\begin{figure*}[tb!]
\begin{verbatim}
grammar ExpressionEvaluator;
expression1 returns [double value]
            : e=expression2 {$value = $e.value;}
              ( '+' e2=expression1 {$value += $e2.value;}
              | '-' e2=expression1 {$value -= $e2.value;}
              )*
            ;
expression2 returns [double value]
            : e=expression3 {$value = $e.value;}
              ( '*' e2=expression2 {$value *= $e2.value;}
              | '/' e2=expression2 {$value -= $e2.value;}
              )*
            ;
expression3 returns [double value]
            : '(' e=expression1 ')' {$value = $e.value;}
            | '+' e=expression1 {$value = $e.value;}
            | '-' e=expression1 {$value = -$e.value;}
            | i=INTEGERLITERAL {$value = (double)Integer.parseInt($i.text);}
            | f=FLOATLITERAL {$value = Double.parseDouble($f.text);}
            ;
INTEGERLITERAL : '0'..'9'+ ;
FLOATLITERAL : ('0'..'9')+ '.' ('0'..'9')* ;
NEWLINE : '\r'? '\n' ;
\end{verbatim}
\caption{Complete ANTLR implementation of the arithmetic expression interpreter.} \label{fig:calcimantlr}
\end{figure*}

Since ANTLR provides no mechanism for the declarative specification of token precedences, such precedences have to be incorporated into the grammar.
The usual solution involves the creation of different nonterminal symbols in the grammar, so that productions corresponding to the same precedence levels are grouped together.
The productions with \emph{expression1} and \emph{expression2} in their left-hand side were introduced with this purpose in our arithmetic expression grammar.

Likewise, since ANTLR generates a LL(*) parser, which does not support left-recursion, left-recursive grammar productions in the grammar shown in Figure \ref{fig:calctrad} have to be refactored.
In our example, a simple solution involves the introduction of the \emph{expression3} nonterminal, which in conjunction with the aforementioned \emph{expression1} and \emph{expression2} nonterminals, eliminates left-recursion from our grammar.

Once the grammar is adjusted to satisfy the constraints imposed by the ANTLR parser generator, the language designer can define the semantic actions needed to implement our arithmetic expression interpreter.
The resulting ANTLR implementation is shown in Figure \ref{fig:calcimantlr}.
The streamlined syntax of the scannerless ANTLR parser generator makes this implementation significantly more concise than the equivalent lex \& yacc implementation.
However, the constraints imposed by the underlying parsing algorithm forces explicit changes on the language grammar (cf. BNF grammar in Figure \ref{fig:calctrad}).

\subsection{ModelCC Implementation}

\begin{figure*}[tb!]
\centering
\includegraphics[scale=1]{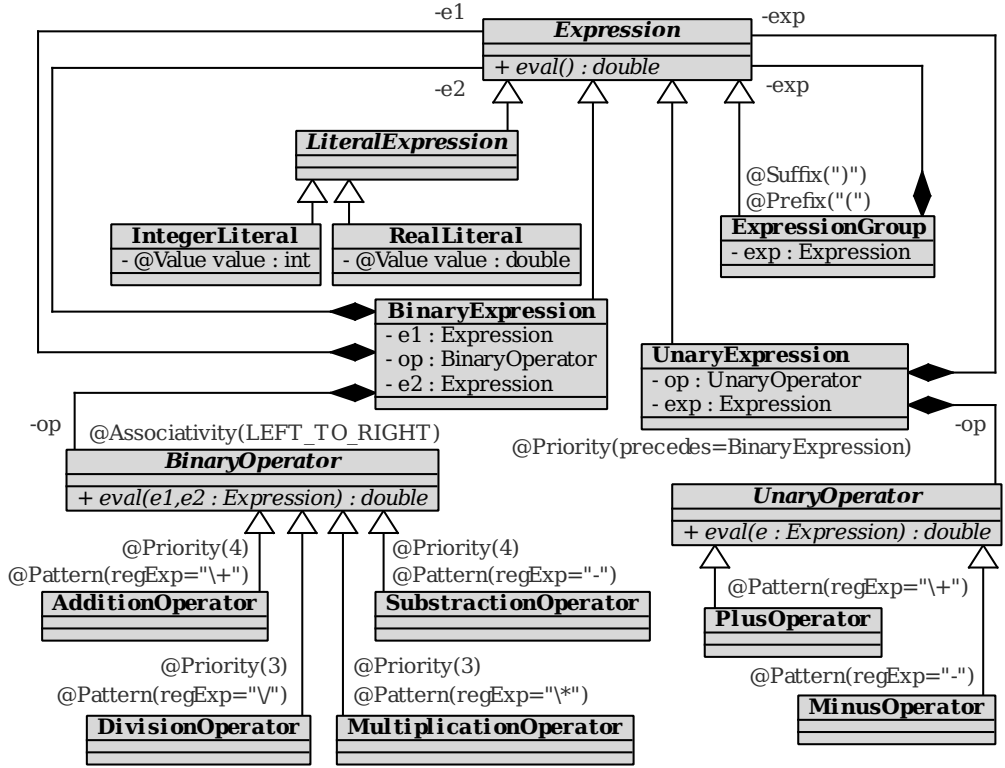}
\caption{ModelCC specification of the arithmetic expression language.} \label{fig:calcmodelcc}
\end{figure*}

When following a model-based language specification approach, the language designer starts by elaborating an abstract syntax model, which will later be mapped to a concrete syntax model by imposing constraints on the abstract syntax model.
These constraints can also be specified as metadata annotations on the abstract syntax model and the resulting annotated model can be processed by automated tools, such as ModelCC, to generate the corresponding lexers and parsers.
Annotated models can be represented graphically, as the UML diagram in Figure \ref{fig:calcmodelcc}, or implemented using conventional programming languages, as the Java implementation listed in Figures \ref{fig:calcimmodelcc1} and \ref{fig:calcimmodelcc2}.

In this example, metadata annotations are used for the ASM-CSM mapping, since a single concrete syntax model is needed.
In case different CSMs were required, the ModelCC domain-specific language for ASM-CSM mappings could be used to specify alternative CSMs for the language ASM \cite{Quesada2014b}.

\begin{figure*}[p!]
\begin{verbatim}
public abstract class Expression implements IModel {
  public abstract double eval();
}
\end{verbatim}
\sepv
\begin{verbatim}
@Prefix("\\(") @Suffix("\\)")
public class GroupExpression extends Expression implements IModel {
  Expression e;
  @Override public double eval() { return e.eval(); }
}
\end{verbatim}
\sepv
\begin{verbatim}
public abstract class LiteralExpression extends Expression implements IModel {
}
\end{verbatim}
\sepv
\begin{verbatim}
public class UnaryExpression extends Expression implements IModel {
  UnaryOperator op;
  Expression e;
  @Override public double eval() { return op.eval(e); }
}
\end{verbatim}
\sepv
\begin{verbatim}
public class BinaryExpression extends Expression implements IModel {
  Expression e1;
  BinaryOperator op;
  Expression e2;
  @Override public double eval() { return op.eval(e1,e2); }
}
\end{verbatim}
\sepv
\begin{verbatim}
public class IntegerLiteral extends LiteralExpression implements IModel {
  @Value int value;
  @Override public double eval() { return (double)value; }
}
\end{verbatim}
\sepv
\begin{verbatim}
public class RealLiteral extends LiteralExpression implements IModel {
  @Value double value;
  @Override public double eval() { return value; }
}
\end{verbatim}
\sepv
\begin{verbatim}
public abstract class UnaryOperator implements IModel {
  public abstract double eval(Expression e);
}
\end{verbatim}
\sepv
\begin{verbatim}
@Pattern(regExp="\\+")
public class PlusOperator extends UnaryOperator implements IModel {
  @Override public double eval(Expression e) { return e.eval(); }
}
\end{verbatim}
\sepv
\begin{verbatim}
@Pattern(regExp="-")
public class MinusOperator extends UnaryOperator implements IModel {
  @Override public double eval(Expression e) { return -e.eval(); }
}
\end{verbatim}
\sepv
\begin{verbatim}
@Associativity(AssociativityType.LEFT_TO_RIGHT)
public abstract class BinaryOperator implements IModel {
  public abstract double eval(Expression e1,Expression e2);
}
\end{verbatim}
\caption{Complete Java implementation of the arithmetic expression interpreter using ModelCC: A set of Java classes define the language ASM,
metadata annotations specify the desired ASM-CSM mapping, and object methods implement arithmetic expression evaluation (1/2).}
\label{fig:calcimmodelcc1}
\end{figure*}
\begin{figure*}[tb!]
\begin{verbatim}
@Priority(value=2) @Pattern(regExp="\\+")
public class AdditionOperator extends BinaryOperator implements IModel {
  @Override public double eval(Expression e1,Expression e2) { return e1.eval()+e2.eval(); }
}
\end{verbatim}
\sepv
\begin{verbatim}
@Priority(value=2) @Pattern(regExp="-")
public class SubtractionOperator extends BinaryOperator implements IModel {
  @Override public double eval(Expression e1,Expression e2) { return e1.eval()-e2.eval(); }
}
\end{verbatim}
\sepv
\begin{verbatim}
@Priority(value=1) @Pattern(regExp="\\*")
public class MultiplicationOperator extends BinaryOperator implements IModel {
  @Override public double eval(Expression e1,Expression e2) { return e1.eval()*e2.eval(); }
}
\end{verbatim}
\sepv
\begin{verbatim}
@Priority(value=1) @Pattern(regExp="\\/")
public class DivisionOperator extends BinaryOperator implements IModel {
  @Override public double eval(Expression e1,Expression e2) { return e1.eval()/e2.eval(); }
}
\end{verbatim}
\caption{Complete Java implementation of the arithmetic expression interpreter using ModelCC: A set of Java classes define the language ASM,
metadata annotations specify the desired ASM-CSM mapping, and object methods implement arithmetic expression evaluation (2/2).}
\label{fig:calcimmodelcc2}
\end{figure*}

The parser that ModelCC generates from the simple arithmetic expression data model can parse input strings such as ``10/(2+3)*0.5+1'' and produce \emph{Expression} objects out of them.
The \emph{eval} method yields the final result for any expression (2, in the previous example).
Figure \ref{fig:run} shows the actual code needed to generate and invoke the parser in ModelCC.

\begin{figure*}[tb!]
\centering
\begin{verbatim}
// Read the model.
Model model = JavaModelReader.read(Expression.class);
// Generate the parser.
Parser<Expression> parser = ParserFactory.create(model);
// Parse a string and directly instantiate the corresponding expression.
Expression expr = parser.parse("10/(2+3)*0.5+1");
// Evaluate the expression.
double value = expr.eval();
\end{verbatim}
\caption{A code snippet showing how the arithmetic expression parser is generated and invoked.} \label{fig:run}
\end{figure*}

ModelCC, in particular, generates Lamb lexers \cite{Quesada2011a} and Fence parsers \cite{Quesada2012f}, albeit traditional LL and LR parsers might also be generated whenever the ASM-CSM mapping constraints make LL and LR parsing feasible.
Whitespace and comments in the language can be defined by specifying ignore patterns during the parser creation.

It should be noted that parse error handling is also completely dependant on the parser being used. Indeed, most parsers are able to provide comprehensive parsing error tracebacks.

However, ModelCC provides a testing framework that integrates well with existing IDEs and JUnit.
Since separate language elements are models themselves, it is possible to implement both unitary and integration tests that focus on specific language elements.
For example, assertions can check whether a model matches a certain string, whether a model does not match a certain string, or whether a model matches a string in a specific number of ways (e.g. matching without ambiguities).
Assertions can, of course, take into consideration the contents or members of language elements for in-depth testing.

Since the abstract syntax model in ModelCC is not constrained by the vagaries of particular parsing algorithms, the language design process can be focused on its conceptual design, without having to introduce artifacts in the design just to satisfy the demands of particular tools:

\begin{itemize}

\item
As we saw in the lex \& yacc example, conventional tools, unless they are scannerless, force the creation of artificial token types in order to avoid lexical ambiguities, which leads to duplicate grammar production rules and semantic actions in the language specification.
As in any other software development project, duplication hinders the evolution of languages and affects the maintainability of language processors.
ModelCC, even though it is not scannerless, supports lexical ambiguities and each basic language element is defined as a separate and independent entity, even when their pattern specification are in conflict.
Therefore, duplication in the language model does not have to be included to deal with lexical ambiguities: token type definitions do not have to be adjusted, duplicate syntactic constructs rules will not appear in the language model, and, as a consequence, semantic predicates do not have to be duplicated either.

\item
As we also saw both in the lex \& yacc interpreter and in the ANTLR solution to the same problem, established parser generators require modifications to the language grammar specification in order to comply with parsing constraints, let it be the elimination of left-recursion for LL parsers or the introduction of new nonterminals to restructure the language specification so that the desired precedence relationships are fulfilled.
In the model-driven language specification approach, the left-recursion problem disappears since it is something the underlying tool can easily deal with in a fully automated way when an abstract syntax model is converted into a concrete syntax model.
Moreover, the declarative specification of constraints, such as the evaluation order constraints in Section \ref{sec:modelconstraints}, is orthogonal to the abstract syntax model that defines the language.
Those constraints determine the ASM-CSM mapping and, since ModelCC takes charge of everything in that conversion process, the language designer does not have to modify the abstract syntax model just because a given parser generator might prefer its input in a particular format.
This is the main benefit that results from raising your abstraction level in model-based language specification.

\item
When changes in the language specification are necessary, as it is often the case when a software system is successful, the traditional language designer will have to propagate changes throughout the entire language processing tool chain, often introducing significant changes and making profound restructurings in the working code base.
The changes can be time-consuming, quite tedious, and extremely error-prone.
In contrast, modifications are more easily done when a model-driven language specification approach is followed.
Any modifications in a language will affect either to the abstract syntax model, when new capabilities are incorporated into a language, or to the constraints that define the ASM-CSM mapping, whenever syntactic details change or new CSMs are devised for the same ASM.
In either case, the more time-consuming, tedious, and error-prone modifications are automated by ModelCC, whereas the language designer can focus his efforts on the essential part of the required changes.

\item
Traditional parser generators typically mix semantic actions with the syntactic details of the language specification.
This approach, which is justified when performance is the top concern, might lead to poorly-designed hard-to-test systems when not done with extreme care.
Moreover, when different applications or tools employ the same language, any changes to the syntax of that language have to be replicated in all the applications and tools that use the language.
The maintenance of several versions of the same language specification in parallel might also lead to severe maintenance problems.
In contrast, the separation of concerns provided by ModelCC, as separate ASM and ASM-CSM mappings, promotes a more elegant design for language processing systems.
By decoupling language specification from language processing and providing a conceptual model for the language, different applications and tools can now use the same language without having duplicate language specifications.
A similar result could be hand-crafted using traditional parser generators (i.e. making their implicit conceptual model explicit and working on that explicit model), but ModelCC automates this part of the process.

\end{itemize}

In summary, while traditional language processing tools provide different mechanisms for resolving ambiguities and implementing language constraints, the solutions they provide typically interfere with the conceptual modeling of languages: relatively minor syntactic details might significantly affect the structure of the whole language specification.
Model-driven language specification, as exemplified by ModelCC, provides a cleaner separation of concerns: the abstract syntax model is kept separate from its incarnation in concrete syntax models, thereby separating the specification of abstractions in the ASM from the particularities of their textual representation in CSMs.

\section{A More Complex Example} \label{sec:example2}

In this section, we present an example imperative arithmetic language that illustrates language composition and reference resolution in ModelCC.

First, we will outline the features we wish to include in our imperative arithmetic language.
Then, we will provide the full language specification for ModelCC by defining an abstract syntax model, which will be annotated to specify the desired ASM-CSM mapping.

\subsection{Language Description}

Our imperative arithmetic language is designed to support the following features:

\begin{itemize}
\item Programs, which consist of a function.
\item Functions, which define scopes, support parameters, and allow the declaration of variables and nested functions.
\item Simple or array variables.
\item Arithmetic expressions as described by the example language in the previous section, with added support for variable expressions, function call expressions, and relational and logical operators.
\item Block statements.
\item Assignment statements.
\item Conditional statements.
\item Repetitive (i.e. while loop) statements.
\item Return statements, for usage in functions.
\item Expression statements, which evaluate an expression (e.g. a function call expression).
\item Predefined input/output functions such as ``read'' and ``print''.
\item Predefined mathematical functions such as ``sin'', ``cos'', ``tan'', ``arcsin'', ``arccos'', ``arctan'', ``floor'', ``ceil'', ``round'', ``power'', ``root'', and ``log''. Modifiable.
\item Predefined variables such as ``pi'' or ``e''. Modifiable.
\end{itemize}

Figure \ref{fig:imparitprog} illustrates an example input of the imperative arithmetic language that defines functions, uses arithmetic expressions, and uses input/output statements.

\begin{figure*}[tb!]
\centering
\begin{verbatim}
function main()
  variables radius,height;
  function diameter(radius)
    return radius*2;
  function perimeter(radius)
    return 2*pi*radius;
  function diskArea(radius)
    return power(pi*radius,2);
  function rectangleArea(side1,side2)
    return side1*side2;
  function cuboidVolume(side1,side2,height)
    return rectangleArea(side1,side2)*height;
  function cylinderVolume(radius,height)
    return diskArea(radius)*height;
  function sphereArea(radius)
    return 4*pi*power(radius,2);
  function sphereVolume(radius)
    return 4/3*pi*power(radius,3);
  begin
    % Read radius and height of a cylinder.
    radius = read();
    height = read();
    % Check for invalid input.
    if (radius<=0 || height<=0) return -1;
    % Print cylinder volume.
    print(cylinderVolume(radius,height));
  end
\end{verbatim}
\caption{An example input of the imperative arithmetic language.} \label{fig:imparitprog}
\end{figure*}

\begin{figure}[p]
\centering
\includegraphics[scale=1]{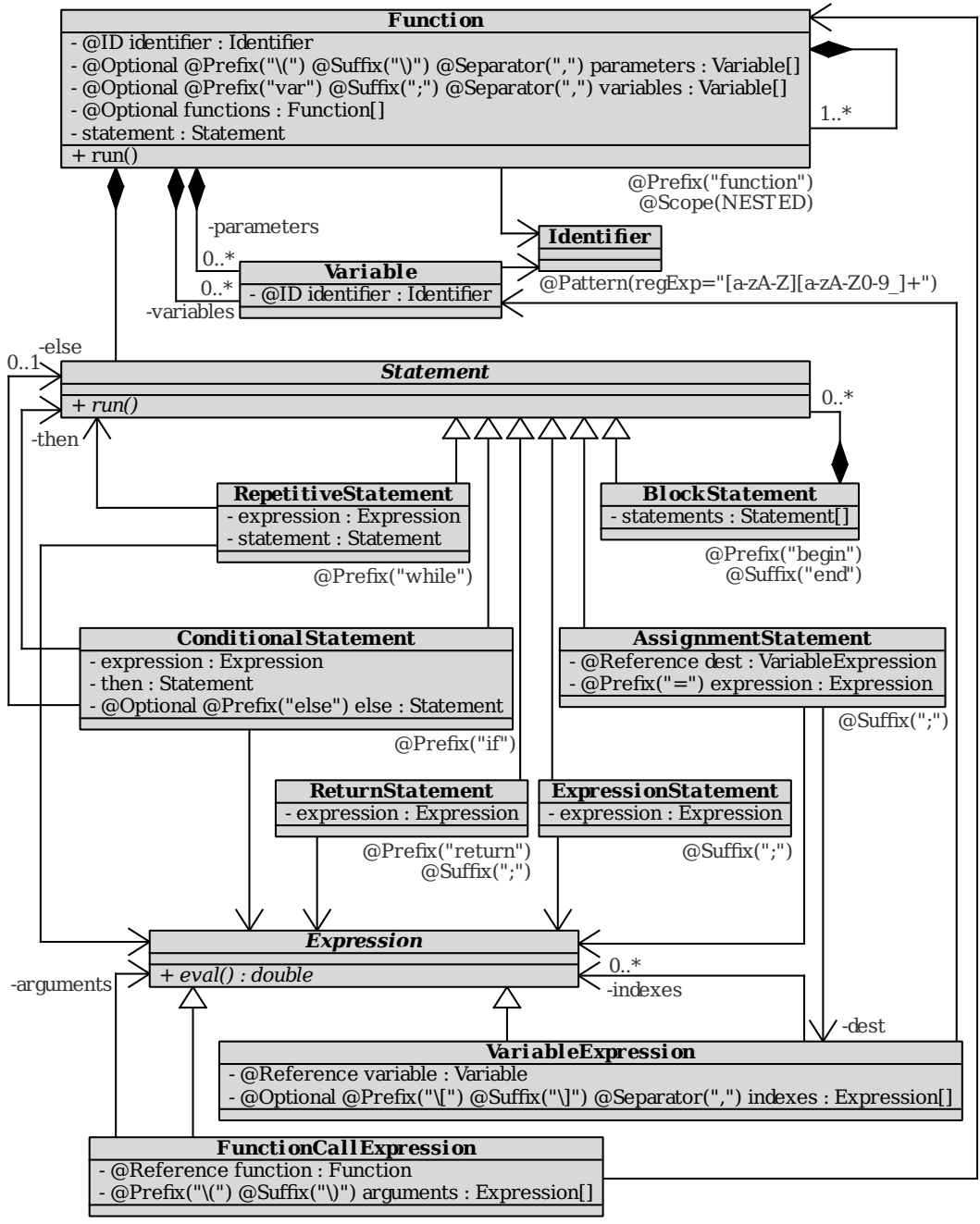}
\caption{ModelCC specification of an imperative arithmetic language. ModelCC reference resolution support is used to allow the declaration of variables and functions. ModelCC language composition support is used to include \emph{Expressions}, which were defined as a separate language in Figure \ref{fig:calcmodelcc}.} \label{fig:imparitlang}
\end{figure}

\begin{figure}[tb]
\centering
\includegraphics[scale=1]{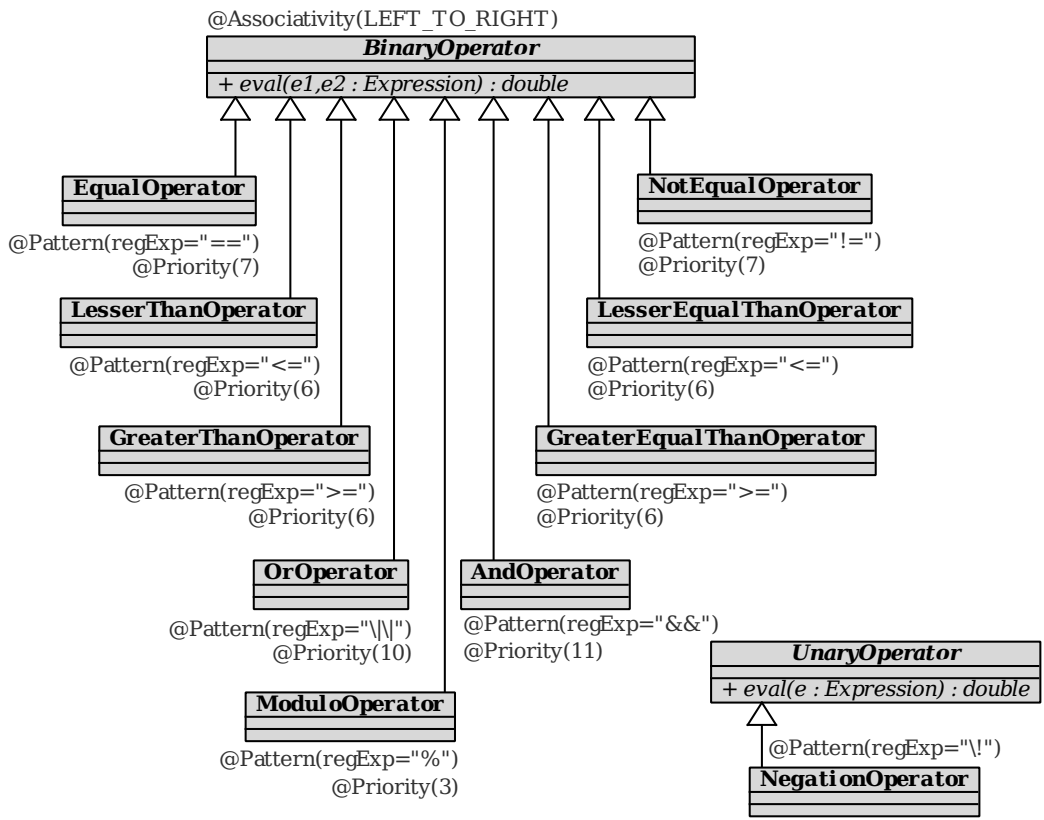}
\caption{ModelCC specification of relational and logical binary operators that extend the simple arithmetic expression language in Figure \ref{fig:calcmodelcc} for the imperative arithmetic language.} \label{fig:imparitlangop}
\end{figure}

\begin{figure}[tb]
\begin{verbatim}
// Read the model.
Model model = JavaModelReader.read(Program.class);
// Generate the parser.
Parser<Program> parser = ParserFactory.create(model);
// Parse a file and directly instantiate the corresponding program.
Program program = parser.parse(new FileInputStream("example.txt"));
// Run the program.
program.run();
\end{verbatim}
\caption{A code snippet showing how the imperative arithmetic language parser is generated and invoked.} \label{fig:run2}
\end{figure}

\subsection{ModelCC Implementation}

In ModelCC, the abstract syntax model is designed first and then it is mapped to a concrete syntax model by imposing constraints by means of metadata annotations on the abstract syntax model.

The resulting model can be processed by ModelCC to generate the corresponding parser.
The UML class diagrams in Figure \ref{fig:imparitlang} presents our annotated imperative arithmetic expression language, which is complemented by the arithmetic expression language in Figure \ref{fig:calcmodelcc} and the binary operators in Figure \ref{fig:imparitlangop}.
Figure \ref{fig:run2} shows the actual code needed to generate and invoke the parser in ModelCC.

Predefined functions and variables are implemented by adding the corresponding objects to a global scope prior to the parsing.

This example illustrates ModelCC capabilities for language composition: the simple arithmetic expression language described in Section \ref{sec:example1} is not only used within the imperative arithmetic language, but it is also extended with new expression types and binary operators.

The reference support extension we propose in this paper can be observed in the \emph{Function} and \emph{Variable} classes.
The \emph{identifier} member of the \emph{Function} and \emph{Variable} classes is annotated with \emph{@ID}, which means that \emph{Function} or \emph{Variable} instances can be identified by an \emph{Identifier}.

Then, the \emph{function} member of a \emph{FunctionCallExpression} and the \emph{variable} member of a \emph{VariableExpression} are annotated with \emph{@Reference}, which means that, in textual form, a \emph{FunctionCallExpression} can refer to a \emph{Function} by its \emph{Identifier} and a \emph{VariableExpression} can refer to a \emph{Variable} by its \emph{Identifier}.

Each function defines a new nested scope.
Nested scopes, apart from reference resolution, allow variable and function shadowing.

ModelCC is able to automatically generate a grammar from the ASM defined by a class model and the ASM-CSM mapping defined as a set of metadata annotations on the class model.
References in that grammar are automatically resolved by ModelCC so that implementing a symbol table is not needed.

The ModelCC website http://www.modelcc.org contains an assortment of more sample languages implemented using ModelCC.

\section{Conclusions and Future Work} \label{sec:conclusionsfuturework}

In this paper, we have introduced ModelCC, a model-based tool for language specification.
ModelCC lets language designers create explicit models of the concepts a language represents, i.e. the abstract syntax model (ASM) of the language.
Then, that abstract syntax can be represented in textual or graphical form, using the concrete syntax defined by a concrete syntax model (CSM).
ModelCC automates the ASM-CSM mapping by means of metadata annotations on the ASM, which let ModelCC act as a model-based parser generator.

ModelCC is not bound to particular scanning and parsing techniques, so that language designers do not have to tweak their models to comply with the constraints imposed by particular parsing algorithms.
ModelCC abstracts away many details traditional language processing tools have to deal with.
It cleanly separates language specification from language processing.
Given the proper ASM-CSM mapping definition, ModelCC-generated parsers are able to automatically instantiate the ASM given an input string representing the ASM in a concrete syntax.

Apart from being able to deal with ambiguous languages, ModelCC also allows the declarative resolution of any language ambiguities by means of constraints defined over the ASM.
The current version of ModelCC also supports lexical ambiguities and custom pattern matching classes.

ModelCC also incorporates reference resolution within the parsing process.
Instead of returning abstract syntax trees, ModelCC is able to obtain abstract syntax graphs from its input string.
Such abstract syntax graphs are not restricted to directed acyclic graphs, since ModelCC supports the resolution of anaphoric, cataphoric, and recursive references.

The proposed model-driven language specification approach promotes the domain-driven design of language processors.
Its model-driven philosophy supports language evolution by improving the maintainability of languages processing system.
It also facilitates the reuse of language specifications across product lines and different applications, eliminating the duplication required by conventional tools and improving the modularity of the resulting systems.

A fully-functional version of ModelCC is available at http://www.modelcc.org.

In the future, we plan to further study the possibilities tools such as ModelCC open up in different application domains, including traditional language processing systems (compilers and interpreters) \cite{Aho2006}, domain-specific languages \cite{Fowler2010,Hudak1996,Mernik2005} and language workbenches \cite{language-workbenches}, model-driven software development tools \cite{Schmidt2006,mdsd-ideal}, natural language processing \cite{Jurafsky2009,Quesada2013b}, text mining applications \cite{mining12}, data integration \cite{doan2012principles}, and information extraction \cite{Sarawagi2008}.

\bibliographystyle{abbrv}
\bibliography{doc}

\end{document}